\documentclass[aps,prl, reprint, superscriptaddress, longbibliography]{revtex4-1}
\usepackage{graphicx}
\usepackage{xcolor}
\usepackage{amsmath}
\usepackage{hyperref}

\hypersetup{
    colorlinks=true,
    linkcolor=blue,
    filecolor=magenta,      
    urlcolor=blue,
    citecolor=blue
}

\usepackage{amssymb}
\usepackage{multirow,array}

\newcommand{\be}{\begin{equation}}
\newcommand{\ee}{\end{equation}}
\newcommand{\bs}{\begin{subequations}}
\newcommand{\es}{\end{subequations}}
\newcommand{\beal}{\begin{align}}

\newcommand{\trento}{\texttt{T$_\mathrm{R}$ENTo }}

\newcommand{\sqrts}{\sqrt{s_\textrm{NN}}}

\begin{document}

\title{Phenomenological constraints on the transport properties of QCD matter\\ with data-driven model averaging}

\date{\today}


\author{D.~Everett}
\affiliation{Department of Physics, The Ohio State University, Columbus OH 43210.}

\author{W.~Ke}
\affiliation{Department of Physics, University of California, Berkeley CA 94270.}
\affiliation{Nuclear Science Division, Lawrence Berkeley National Laboratory, Berkeley CA 94270.}

\author{J.-F. Paquet}
\affiliation{Department of Physics, Duke University, Durham NC 27708.}

\author{G.~Vujanovic}
\affiliation{Department of Physics and Astronomy, Wayne State University, Detroit MI 48201.}

\author{S.~A.~Bass}
\affiliation{Department of Physics, Duke University, Durham NC 27708.}

\author{L.~Du}
\affiliation{Department of Physics, The Ohio State University, Columbus OH 43210.}

\author{C.~Gale}
\affiliation{Department of Physics and Astronomy, McGill University, Montr\'{e}al QC H3A-2T8.}

\author{M.~Heffernan}
\affiliation{Department of Physics and Astronomy, McGill University, Montr\'{e}al QC H3A-2T8.}

\author{U.~Heinz}
\affiliation{Department of Physics, The Ohio State University, Columbus OH 43210.}

\author{D.~Liyanage}
\affiliation{Department of Physics, The Ohio State University, Columbus OH 43210.}

\author{M.~Luzum}
\affiliation{Instituto  de  F\`{i}sica,  Universidade  de  S\~{a}o  Paulo,  C.P.  66318,  05315-970  S\~{a}o  Paulo,  SP,  Brazil. }

\author{A.~Majumder}
\affiliation{Department of Physics and Astronomy, Wayne State University, Detroit MI 48201.}

\author{M.~McNelis}
\affiliation{Department of Physics, The Ohio State University, Columbus OH 43210.}

\author{C.~Shen}
\affiliation{Department of Physics and Astronomy, Wayne State University, Detroit MI 48201.}
\affiliation{RIKEN BNL Research Center, Brookhaven National Laboratory, Upton NY 11973.}

\author{Y.~Xu}
\affiliation{Department of Physics, Duke University, Durham NC 27708.}


\author{A.~Angerami}
\affiliation{Lawrence Livermore National Laboratory, Livermore CA 94550.}

\author{S.~Cao}
\affiliation{Department of Physics and Astronomy, Wayne State University, Detroit MI 48201.}

\author{Y.~Chen}
\affiliation{Laboratory for Nuclear Science, Massachusetts Institute of Technology, Cambridge MA 02139.}
\affiliation{Department of Physics, Massachusetts Institute of Technology, Cambridge MA 02139.}

\author{J.~Coleman}
\affiliation{Department of Statistical Science, Duke University, Durham NC 27708.}

\author{L.~Cunqueiro}
\affiliation{Department of Physics and Astronomy, University of Tennessee, Knoxville TN 37996.}
\affiliation{Physics Division, Oak Ridge National Laboratory, Oak Ridge TN 37830.}

\author{T.~Dai}
\affiliation{Department of Physics, Duke University, Durham NC 27708.}

\author{R.~Ehlers}
\affiliation{Department of Physics and Astronomy, University of Tennessee, Knoxville TN 37996.}
\affiliation{Physics Division, Oak Ridge National Laboratory, Oak Ridge TN 37830.}

\author{H.~Elfner}
\affiliation{GSI Helmholtzzentrum f\"{u}r Schwerionenforschung, 64291 Darmstadt, Germany.}
\affiliation{Institute for Theoretical Physics, Goethe University, 60438 Frankfurt am Main, Germany.}
\affiliation{Frankfurt Institute for Advanced Studies, 60438 Frankfurt am Main, Germany.}

\author{W.~Fan}
\affiliation{Department of Physics, Duke University, Durham NC 27708.}

\author{R.~J.~Fries}
\affiliation{Cyclotron Institute, Texas A\&M University, College Station TX 77843.}
\affiliation{Department of Physics and Astronomy, Texas A\&M University, College Station TX 77843.}

\author{F.~Garza}
\affiliation{Cyclotron Institute, Texas A\&M University, College Station TX 77843.}
\affiliation{Department of Physics and Astronomy, Texas A\&M University, College Station TX 77843.}

\author{Y.~He}
\affiliation{Key Laboratory of Quark and Lepton Physics (MOE) and Institute of Particle Physics, Central China Normal University, Wuhan 430079, China.}

\author{B.~V.~Jacak}
\affiliation{Department of Physics, University of California, Berkeley CA 94270.}
\affiliation{Nuclear Science Division, Lawrence Berkeley National Laboratory, Berkeley CA 94270.}

\author{P.~M.~Jacobs}
\affiliation{Department of Physics, University of California, Berkeley CA 94270.}
\affiliation{Nuclear Science Division, Lawrence Berkeley National Laboratory, Berkeley CA 94270.}

\author{S.~Jeon}
\affiliation{Department of Physics and Astronomy, McGill University, Montr\'{e}al QC H3A-2T8.}



\author{B.~Kim}
\affiliation{Cyclotron Institute, Texas A\&M University, College Station TX 77843.}
\affiliation{Department of Physics and Astronomy, Texas A\&M University, College Station TX 77843.}

\author{M.~Kordell~II}
\affiliation{Cyclotron Institute, Texas A\&M University, College Station TX 77843.}
\affiliation{Department of Physics and Astronomy, Texas A\&M University, College Station TX 77843.}

\author{A.~Kumar}
\affiliation{Department of Physics and Astronomy, Wayne State University, Detroit MI 48201.}


\author{S.~Mak}
\affiliation{Department of Statistics, Duke University, Durham NC 27708.}

\author{J.~Mulligan}
\affiliation{Department of Physics, University of California, Berkeley CA 94270.}
\affiliation{Nuclear Science Division, Lawrence Berkeley National Laboratory, Berkeley CA 94270.}

\author{C.~Nattrass}
\affiliation{Department of Physics and Astronomy, University of Tennessee, Knoxville TN 37996.}

\author{D.~Oliinychenko}
\affiliation{Nuclear Science Division, Lawrence Berkeley National Laboratory, Berkeley CA 94270.}



\author{C. Park}
\affiliation{Department of Physics and Astronomy, McGill University, Montr\'{e}al QC H3A-2T8.}

\author{J.~H.~Putschke}
\affiliation{Department of Physics and Astronomy, Wayne State University, Detroit MI 48201.}

\author{G.~Roland}
\affiliation{Laboratory for Nuclear Science, Massachusetts Institute of Technology, Cambridge MA 02139.}
\affiliation{Department of Physics, Massachusetts Institute of Technology, Cambridge MA 02139.}

\author{B.~Schenke}
\affiliation{Physics Department, Brookhaven National Laboratory, Upton NY 11973.}

\author{L.~Schwiebert}
\affiliation{Department of Computer Science, Wayne State University, Detroit MI 48202.}

\author{A.~Silva}
\affiliation{Department of Physics and Astronomy, University of Tennessee, Knoxville TN 37996.}

\author{C.~Sirimanna}
\affiliation{Department of Physics and Astronomy, Wayne State University, Detroit MI 48201.}

\author{R.~A.~Soltz}
\affiliation{Department of Physics and Astronomy, Wayne State University, Detroit MI 48201.}
\affiliation{Lawrence Livermore National Laboratory, Livermore CA 94550.}

\author{Y.~Tachibana}
\affiliation{Department of Physics and Astronomy, Wayne State University, Detroit MI 48201.}

\author{X.-N.~Wang}
\affiliation{Key Laboratory of Quark and Lepton Physics (MOE) and Institute of Particle Physics, Central China Normal University, Wuhan 430079, China.}
\affiliation{Department of Physics, University of California, Berkeley CA 94270.}
\affiliation{Nuclear Science Division, Lawrence Berkeley National Laboratory, Berkeley CA 94270.}

\author{R.~L.~Wolpert}
\affiliation{Department of Statistical Science, Duke University, Durham NC 27708.}


\collaboration{The JETSCAPE Collaboration}

\begin{abstract}
Using combined data from the Relativistic Heavy Ion and Large Hadron Colliders, we constrain the shear and bulk viscosities of quark-gluon plasma (QGP) at temperatures of ${\sim\,}150{-}350$\,MeV. We use Bayesian inference to translate experimental and theoretical uncertainties into probabilistic constraints for the viscosities. With Bayesian Model Averaging we account for the irreducible model ambiguities in the transition from a fluid description of the QGP to hadronic transport in the final evolution stage, providing the most reliable phenomenological constraints to date on the QGP viscosities.
\end{abstract}

\maketitle

\noindent \textbf{\textit{Introduction.}}
%
Ultrarelativistic collisions of heavy nuclei provide an experimental avenue to produce quark-gluon plasma (QGP), a short-lived state of deconfined hot and dense nuclear matter~\cite{Arsene:2004fa, Back:2004je, Adams:2005dq, Adcox:2004mh, Muller:2012zq, Heinz:2013th, Busza:2018rrf}. The QGP produced in heavy-ion collisions is a strongly-coupled fluid~\cite{Shuryak:2014zxa} that can be characterized by its macroscopic properties such as its equation of state and transport coefficients. These macroscopic characteristics encode the underlying microscopic interactions, described by quantum chromodynamics (QCD), among the fluid's constituents. 

Understanding the different phases of QCD matter remains an area of topical interest in nuclear physics. The equation of state of deconfined nuclear matter with no net baryon density has been known from first principles for more than a decade, by computing the QCD equilibrium partition function numerically on a space-time lattice \cite{Borsanyi:2013bia, Bazavov:2014pvz}. Calculating the transport coefficients of deconfined nuclear plasma, on the other hand, is a continuing challenge: numerical and theoretical uncertainties currently limit the evaluation of the relevant energy-momentum tensor correlators with lattice techniques \cite{Bazavov:2019lgz}; moreover the strongly-coupled microscopic dynamics of the plasma in the experimentally accessible temperature range, ${\sim\,} 150{-}350$\,MeV, largely preclude access via perturbative approaches \cite{Ghiglieri:2018dib, Shuryak:2019ydl}. In this work, we use a large set of hadronic measurements from the Relativistic Heavy Ion Collider (RHIC) and the Large Hadron Collider (LHC) to constrain the temperature dependence of two of the QGP transport coefficients: the shear and bulk viscosities. We improve on earlier such studies \cite{Petersen:2010zt, Novak:2013bqa, Sangaline:2015isa, Bernhard:2015hxa, Bernhard:2016tnd, Bernhard:2019bmu, Yang:2020oig} by accounting for known theoretical ambiguities in our quantification of the uncertainties of the inferred viscosities.

The QGP viscosities have measurable effects on the momentum distribution of hadrons produced in heavy-ion collisions \cite{Romatschke:2007mq, Song:2007fn, Song:2010mg, Heinz:2013th}. A large bulk viscosity isotropically reduces the momenta of the produced hadrons. Shear viscosity, on the other hand, reduces asymmetries in the hadron momentum distributions. Such momentum asymmetries are caused by anisotropic pressure gradients in the QGP, induced by spatial inhomogeneities arising from geometrical and quantum fluctuations at nuclear impact.

The impact of viscosities on the measured hadron distributions is qualitatively well understood. However, their precise effects depend on (among other details) the initial conditions and early (``pre-hydrodynamic'') evolution stage of the collisions, whose quantitative description continues to be challenging \cite{Schlichting:2019abc, Berges:2020fwq}. A further complication is that the effect of different transport coefficients often cannot be factorized. Moreover, due to the dynamics of the plasma, different temperature dependencies of the viscosities may conspire to produce similar effects in certain hadronic observables \cite{Niemi:2015qia, Paquet:2019npk}.

To meaningfully constrain the QGP viscosities from collider measurements, multiple phases of the collision must therefore be precisely modeled and the resulting model predictions compared with large and diverse sets of experimental data. In this work we emphasize that even parts of the collision modeling that appear unrelated to the QGP viscosities can affect their estimation significantly. In particular, we quantify the uncertainties caused by irreducible modeling ambiguities in the transition from a fluid dynamical description of the QGP phase to a microscopic kinetic evolution of the late hadronic phase. With Bayesian inference \cite{von2011bayesian}, and more specifically Bayesian Model Averaging \cite{fragoso2018bayesian}, we account --- in a statistically rigorous fashion --- for both the known experimental and theoretical uncertainties in our probabilistic constraints on the QGP viscosities.

\noindent \textbf{\textit{Modeling heavy-ion collisions.}}
%
In a collision between two heavy ions a fraction of their large kinetic energy is converted into a color-deconfined form of excited nuclear matter. The creation and subsequent evolution of this newly created matter span several different successive many-body regimes of QCD, which can be described with a multistage model. We provide a brief summary of this model, referring for more details to a longer companion paper \cite{long_companion_paper}. The energy deposition during the primary impact is described with the \trento ansatz \cite{Moreland:2014oya, trento_code}; this model accounts for the varying degree of overlap between the colliding nuclei, and for fluctuations in the positions of their nucleons and the amount of energy deposited in each nucleon-nucleon collision. The energy-momentum tensor describing this early, extremely dense stage of the collision is subsequently evolved for a brief period of $\mathcal{O}(0.1{-}1)$\,fm/$c$ as an ensemble of free-streaming massless particles \cite{Liu:2015nwa, Broniowski:2008qk, fs_code}; for sufficiently weakly-coupled systems, it is well established \cite{Vredevoogd:2008id, Keegan:2016cpi, Kurkela:2018vqr, Schlichting:2019abc} that free-streaming approximates well the early pre-hydrodynamic evolution stage. At the end of the free-streaming stage the energy-momentum tensor is matched to (2+1)D dissipative hydrodynamics. In this matching process, space-momentum correlations that developed during the free-streaming stage manifest themselves as non-zero initial flow velocity and viscous stress profiles for the subsequent hydrodynamic evolution. The second-order dissipative relativistic fluid dynamic stage \cite{Schenke:2010nt, Schenke:2010rr, Paquet:2015lta, hydro_code} describes the evolution of the quark-gluon plasma fluid and forms the core of the simulation. Its most important ingredients are a first-principles equation of state from lattice QCD \cite{Bazavov:2014pvz, Bernhard:2018hnz, eos_code} and two parametrized \cite{long_companion_paper} first-order transport coefficients: the specific shear and bulk viscosities. As discussed in the introduction, first-principles information about these viscosities is still limited, and they are here constrained by experimental data. When the fluid has cooled to the pseudo-critical temperature $T_c$, the color-charged QGP constituents recombine into color-neutral hadrons. This color neutralization causes a rapid increase in mean free path, quickly leading to a breakdown of local thermal equilibrium; this requires transitioning \cite{Cooper:1974qi, McNelis:2019auj, is3d_code} from a macroscopic fluid dynamical picture to hadronic kinetic transport \cite{Weil:2016zrk, smash_code, dmytro_oliinychenko_2019_3485108}.

A key aspect of our description of heavy-ion collisions is its unprecedented flexibility. It models a multitude of dynamical details that are known or expected to be phenomenologically important but are theoretically not yet well constrained. For example, we allow the initialization time of hydrodynamics to vary with the amount of energy deposition in the collision, since larger energy densities result in shorter mean free paths and faster hydrodynamization of the fluid \cite{Basar:2013hea}. Combined with the flexibility of the \trento ansatz for the energy deposition followed by free-streaming, this offers a wide range of scenarios for the pre-hydrodynamic collision stage. For the shear and bulk viscosities we employ more general parametrizations than in previous studies \cite{Bernhard:2019bmu}. For example, we assume as in Ref.~\cite{Bernhard:2019bmu} that $\zeta/s(T)$ has a peak in the deconfinement region, but we allow the profile around the peak to be asymmetric in temperature, and we also explore a wider range of values for the width, maximum and position of this peak \cite{long_companion_paper}. We also vary a second-order transport coefficient --- the shear relaxation time --- to quantify its effects on the viscosity constraints.

Most importantly, we extend the previously used Bayesian inference framework \cite{Bernhard:2016tnd, Bernhard:2019bmu} to include the uncertainties resulting from discrete model choices. It is well known \cite{Molnar:2014fva} that the transition from hydrodynamics to hadronic kinetics for the final evolution stage (called {\it particlization}) is an ill-defined problem. The kinetic theory requires initial conditions for the entire hadronic phase-space distribution whereas hydrodynamics provides (in our case) only information about the 10 hydrodynamic moments of the hadronic distribution functions, summed over all hadron species, which make up the energy-momentum tensor $T^{\mu\nu}$. Without additional knowledge about the microscopic dynamics that goes beyond fluid dynamics, this results in an {\it irreducible} uncertainty in the hadron phase-space distribution at the fluid-hadron switching surface. We selected three commonly-used models of the hadronic phase-space distributions in terms of the $T^{\mu\nu}$ components (labeled by $\mathcal{M}_i$, $i{\,=\,}1,2,3$): the 14-moment Grad \cite{CPA:CPA3160020403, Israel:1976tn, Israel:1979wp, Monnai:2009ad, Denicol:2012cn}, relativistic Chapman-Enskog in the relaxation-time approximation (``RTA Chapman-Enskog'') \cite{chapman1990mathematical, Anderson_Witting_1974}, and Pratt-Torrieri-Bernhard \cite{Pratt:2010jt, Bernhard:2018hnz} models. By quantifying, for the first time, the effect of this discrete model ambiguity on the posterior probability distributions, we obtain more reliable constraints on the QGP viscosities than achieved before.

\noindent \textbf{\textit{Data selection.}}
%
We calibrate our model with measurements from the LHC and RHIC. For Pb-Pb collisions at $\sqrts=2.76$\,TeV we use (i) the average number $dN/dy$ of pions, kaons and protons produced in the collisions, along with their average momentum $\langle p_T \rangle$ \cite{Abelev:2013vea}; (ii) the total number of charged hadrons $dN_{\text{ch}}/d\eta$ \cite{Aamodt:2010cz}, along the fluctuation $\delta p_T /\langle p_T\rangle$ of the average momentum \cite{Abelev:2014ckr}; (iii) the total transverse energy of hadrons $dE_T/d\eta$ \cite{Adam:2016thv}; and (iv) the momentum anisotropies $v_n\{2\}$, $n=2,3,4$, of charged hadrons in the plane transverse to the collision axis, as measured through two-particle correlations \cite{ALICE:2011ab}. We further simulate collisions of Au nuclei with $\sqrts=0.2$\,TeV and compare them with a smaller subset of RHIC measurements: $dN/dy$ and $\langle p_T \rangle $ of pions and kaons \cite{Abelev:2008ab}, as well as the momentum anisotropies $v_{2/3}\{2\}$ of charged hadrons \cite{Adams:2004bi, Adamczyk:2013waa}. We note that proton observables are included \emph{only} at the LHC.

\noindent \textbf{\textit{Data-driven constraints on the QGP viscosities.}}
%
A given set of data with uncertainties can be consistent with a range of temperature dependences of the QGP viscosities. Experimental measurements are effectively probability distributions, which we assume normally distributed around their means \footnote{%
    Systematic uncertainties can in principle be non-normal, but generalizing this assumption remains a challenge.}.
The model calculations have uncertainties, too, due to stochastic elements in the fluctuating initial conditions and in the probabilistic sampling of hadrons during particlization. Accordingly, our model predictions are probability distributions as well, with a mean and a standard deviation. 

Let $\mathbf{y}_{\exp}$ denote the full set of experimental measurements and $\boldsymbol{x}$ represent all the model parameters, including those governing the temperature dependences of the QGP viscosities. Discrete model choices are labeled with the index $i$; in our case, this index distinguishes between different particlization models $\mathcal{M}_i$. The model and experimental probability distributions are connected by the ``likelihood'', the probability of the model $\mathcal{M}_i$ being consistent with the data $\mathbf{y}_{\exp}$ at a given value of its parameters $\boldsymbol{x}$:
\begin{equation}
    \mathcal{P}^{(i)}(\mathbf{y}_{\exp}|\boldsymbol{x})
    = \frac{\exp\left(-\frac{\left(\Delta \mathbf{y}_{\boldsymbol{x}}^{(i)}\right)^{\rm T} \Sigma^{-1}_{(i)}(\boldsymbol{x})\Delta \mathbf{y}^{(i)}_{\boldsymbol{x}}}{2} \right)}{\sqrt{(2\pi)^{n} \det\left(\Sigma_{(i)}(\boldsymbol{x})\right)}} .
    \label{eq:likelihood}
\end{equation}
Here $\Delta \mathbf{y}^{(i)}_{\boldsymbol{x}} \equiv\mathbf{y}^{(i)}_{\boldsymbol{x}}{-}\mathbf{y}_{\exp}$ is the discrepancy between the measurements $\mathbf{y}_{\exp}$ and their predictions $\mathbf{y}^{(i)}_{\boldsymbol{x}}$ by model $i$ with the parameters $\boldsymbol{x}$; $n$ is the number of data points (length of $\mathbf{y}_{\exp}$); and $\Sigma_{(i)}(\boldsymbol{x})$ is a covariance matrix encoding both experimental and model uncertainties and their correlations. The experimental contribution to $\Sigma_{(i)}(\boldsymbol{x})$ should in principle be non-diagonal --- we expect many systematic uncertainties to be correlated between data points. As quantitative information on these error correlations is currently limited, the experimental covariance matrix is here assumed to be diagonal.

Constraints on the viscosity are given by the inverse probability $\mathcal{P}^{(i)}(\boldsymbol{x}|\mathbf{y}_{\exp})$ (called ``posterior'') for the parameters $\boldsymbol{x}$ given a set of measurements $\mathbf{y}_{\exp}$, which according to Bayes' theorem is
\begin{equation}
	\mathcal{P}^{(i)}(\boldsymbol{x}|\mathbf{y}_{\exp}) = \frac{\mathcal{P}^{(i)}(\mathbf{y}_{\exp}|\boldsymbol{x}) \mathcal{P}(\boldsymbol{x}) }{\mathcal{P}^{(i)}(\mathbf{y}_{\exp})}.
\label{eq:bayes}
\end{equation}
The dependence of the posterior on $\boldsymbol{x}$ is controlled by two factors: (i) the likelihood $\mathcal{P}^{(i)}(\mathbf{y}_{\exp}|\boldsymbol{x})$ which accounts for the information provided by the measurements $\mathbf{y}_{\exp}$ and their uncertainties, and (ii) the prior probability $\mathcal{P}(\boldsymbol{x})$ that we assign to the parameters {\it before} taking the current data set into account. The prior $\mathcal{P}(\boldsymbol{x})$ reflects our combined theoretical and experimental prior knowledge about the parameters $\boldsymbol{x}$; for example, it allows theoretical constraints on parameters, such as positivity and causality, to be enforced. The normalization of the posterior is controlled by the ``Bayesian model evidence'' $\mathcal{P}^{(i)}(\mathbf{y}_{\exp}) = \int d\boldsymbol{x}\, \mathcal{P}^{(i)} (\mathbf{y}_{\exp}|\boldsymbol{x}) \mathcal{P} (\boldsymbol{x})$, which describes the validity of model $i$ given the data $\mathbf{y}_{\exp}$; as will be discussed below, it can be used to discriminate between different models.

Equation~(\ref{eq:bayes}) shows that the parameter constraints encoded in the posterior $\mathcal{P}^{(i)}(\boldsymbol{x}|\mathbf{y}_{\exp})$ can be improved in two ways: (i) within a given model,
theoretical work can provide tighter constraints for its parameters  $\boldsymbol{x}$, reflected in a tighter prior distribution $\mathcal{P}(\boldsymbol{x})$ \footnote{%
    Theoretical progress can also lead to improvements in the structure of the model itself, 
    not just in its parameter range.},
and (ii) new or more precise experimental data can tighten the likelihood $\mathcal{P}^{(i)}(\mathbf{y}_{\exp}|\boldsymbol{x})$. The ability to include both theoretical and experimental progress consistently and equitably in the extraction of new knowledge is a key feature of Bayesian inference.

\begin{figure}[tb]
\includegraphics[width=\linewidth]{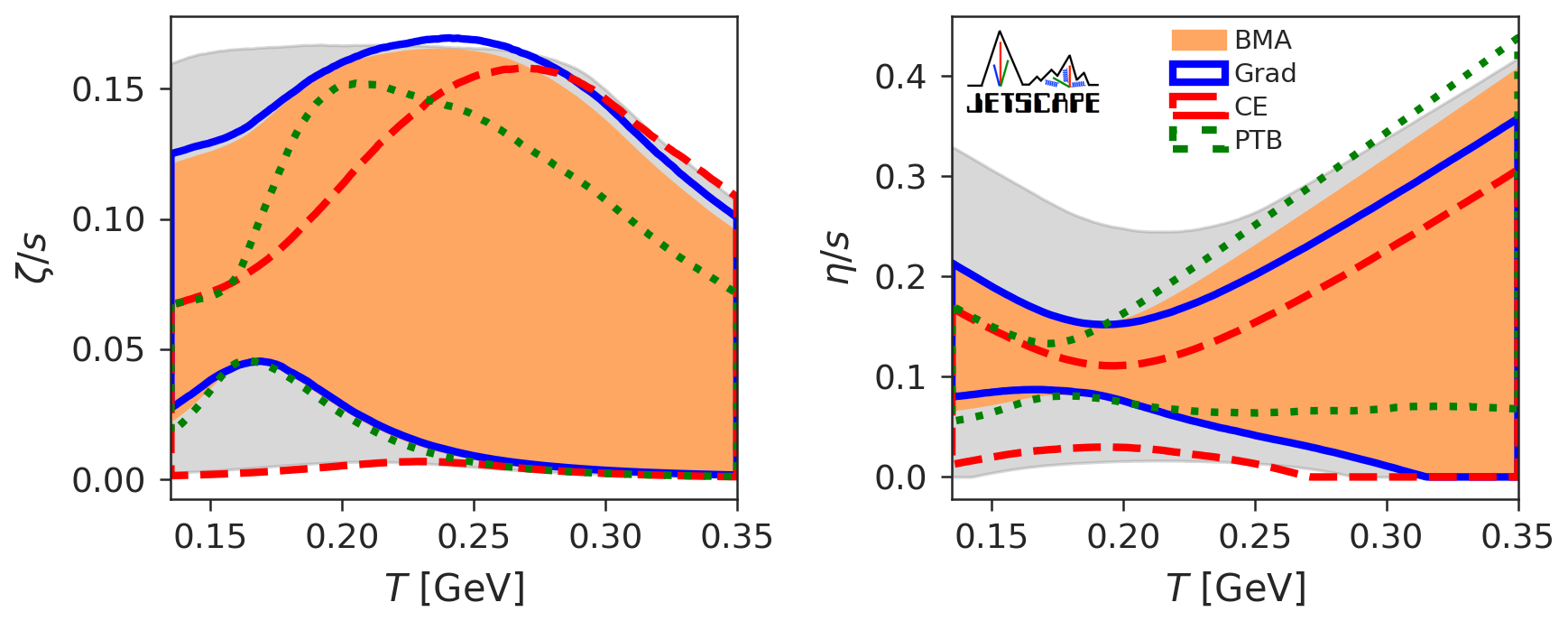}
\caption{The $90$\% credible intervals for the prior (gray), the posteriors of the Grad (blue), Chapman-Enskog (red) and Pratt-Torrieri-Bernhard (green) models, and their Bayesian model average (orange) for the specific bulk (left) and shear (right) viscosities of QGP.
}
\label{F1}
\end{figure}

The temperature dependence of the QGP viscosities favored by the RHIC and LHC data are given by evaluating the posterior (\ref{eq:bayes}) and marginalizing over all parameters except the viscosities. Fig.~\ref{F1} shows the 90\% credibility ranges (outlined by colored lines) for the marginalized posterior of the three particlization models studied here. The high-credibility ranges for the different particlization models show similar qualitative features; however they differ significantly in detail, especially in the low-temperature region between 150 and 250\,MeV where the Bayesian constraints tighten. Importantly, at high temperature, the posteriors are close to the 90\% credibility ranges of the prior (gray shaded region): this strongly suggests that measurements used in this work do not constrain the viscosities significantly for temperatures~$\gtrsim 250$\,MeV. At these high temperatures our results appear to differ significantly from previous works such as Ref.~\cite{Bernhard:2019bmu}. This is mainly a consequence of the choice of prior, $\mathcal{P}(\boldsymbol{x})$ in Eq.~(\ref{eq:bayes}), which can be a double-edged sword: strongly informed priors can overwhelm the constraining power of the data-driven likelihood $\mathcal{P}^{(i)} (\mathbf{y}_{\exp} |\boldsymbol{x})$. This is a benefit if it excludes values of parameters that are considered unlikely on the basis of external evidence; however, it also ties the results of the Bayesian inference to the validity of these additional assumptions. We found \cite{long_companion_paper} that the apparent tighter posterior constraints in Ref.~\cite{Bernhard:2019bmu} are a consequence of their use of narrow priors, and not constraints from measurements. In this sense, the current results are consistent with those of Ref.~\cite{Bernhard:2019bmu}: constraints on the viscosities at high temperatures originate primarily from priors, and not from the data. This conclusion can be easily missed without a careful comparison of posteriors and priors. Exploring the sensitivity of conclusions to prior assumptions must therefore be a key component of future studies.

There is insufficient theoretical evidence at the moment to establish which particlization model is a better description of the process in heavy-ion collisions. In absence of such prior theoretical insights, we use experimental measurements to judge the quality of each particlization model. This is done by using the Bayes evidence $\mathcal{P}^{(i)}(\mathbf{y}_{\exp})$ from Eq.~(\ref{eq:bayes}), which corresponds to the average of the likelihood over the parameter space. The Bayes evidence favors good agreement with data (high likelihood) while disfavoring model complexity, as additional model parameters that do not significantly improve agreement with the data dilute the average of the likelihood \cite{Liddle:2007fy}. The ratio of Bayes evidences is approximately 5000:2000:1 for the Grad, Pratt-Torrieri-Bernhard and Chapman-Enskog particlization models respectively, clearly disfavoring the Chapman-Enskog model.

The Bayesian evidence can be used as a data-driven approach to combine the results for the three particlization models into one posterior distribution \footnote{%
    For simplicity and in accord with the current analysis, we take the models to share a common set of parameters and priors.}, 
as defined by Bayesian Model Averaging \cite{fragoso2018bayesian}:
\begin{equation}
    \mathcal{P}_{\textrm{BMA}}(\boldsymbol{x}|\mathbf{y}_{\exp}) \propto \sum_i \mathcal{P}^{(i)}(\mathbf{y}_{\exp}) \mathcal{P}^{(i)}(\boldsymbol{x}|\mathbf{y}_{\exp}).
    \label{eq:bam}
\end{equation}
This results in the orange band in Fig.~\ref{F1}. Being strongly disfavored by the Bayesian evidence, the impact of the Chapman-Enskog particlization model on the Bayesian model average (\ref{eq:bam}) is minor. 

\begin{figure}[tb]
\includegraphics[width=\linewidth]{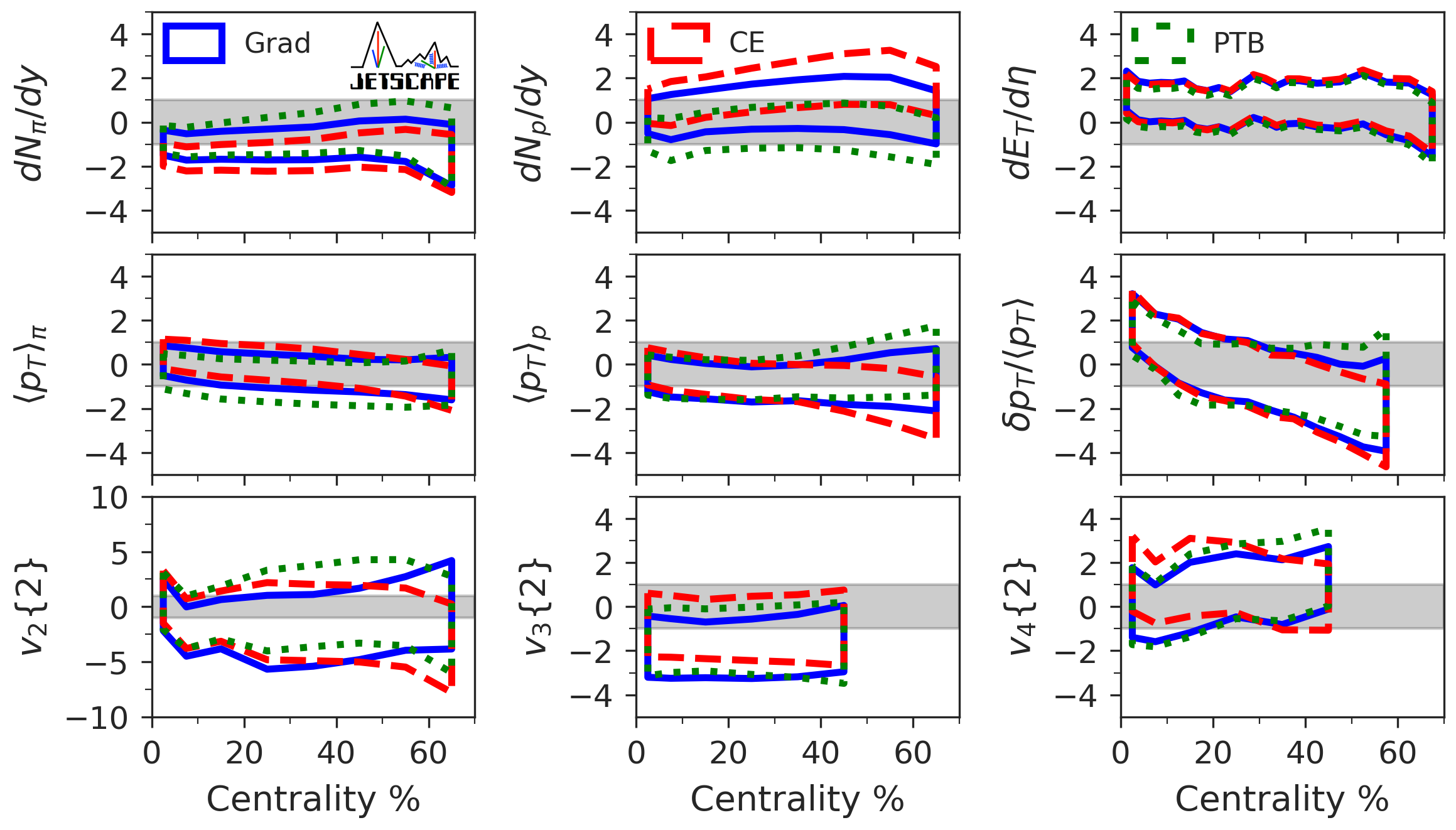}
\caption{The $90$\% credible intervals of the posterior predictive distribution of observables for Pb-Pb collisions at the LHC as functions of centrality, for the Grad (blue), Chapman-Enskog (red) and Pratt-Torrieri-Bernhard (green) particlization models. Plotted is the model discrepancy in units of the experimental standard deviation $\sigma_{\rm exp}$; the vertical axes are labeled with shorthand notation $y \equiv (y_{\rm model}{-}y_{\rm exp}) / \sigma_{\rm exp}$ where $y$ stands for the observable whose model discrepancy is shown. The gray bands represent a discrepancy of one $\sigma_{\rm exp}$ above and below zero. 
\label{F2}
}
\end{figure}

The level of agreement of each particlization model with a representative subset of measurements is shown in Fig.~\ref{F2}. The bands represent the 90\% posterior predictive distributions of observables, obtained by sampling the parameter posterior $\mathcal{P}^{(i)}(\boldsymbol{x}|\mathbf{y}_{\exp})$. All three particlization models show reasonable agreement with the data, giving credence to their respective posterior estimates of the shear and bulk viscosity (and other model parameters) that were inferred from the model-to-data comparison. A closer look at Fig.~\ref{F2} reveals tension with the Chapman-Enskog particlization model, which struggles at describing the pion and proton multiplicities simultaneously. This tension in the proton-to-pion ratio is the origin of its small Bayes evidence. In Ref.~\cite{long_companion_paper} we show that ignoring the proton $dN/dy$ reduces the odds against the Chapman-Enskog particlization model from 5000:1 to 5:1; the key feature behind its failure is the form of its bulk viscous correction to the particle momentum distributions. This highlights the importance of understanding how energy and momentum are distributed across both momentum \emph{and} species at particlization. We note that our choice of likelihood function, Eq.~(\ref{eq:likelihood}), assumes that probability decreases rapidly away from the mean; this can be unforgiving to tension with the data, resulting in the large ratios of Bayes evidence  encountered in this work. Other forms of likelihood should be investigated in the future. Nevertheless, we believe the proton-to-pion ratio is an important observable: the averaged constraints consequently favor particlization models that can describe it well.

\begin{figure}[ht]
\includegraphics[width=\linewidth]{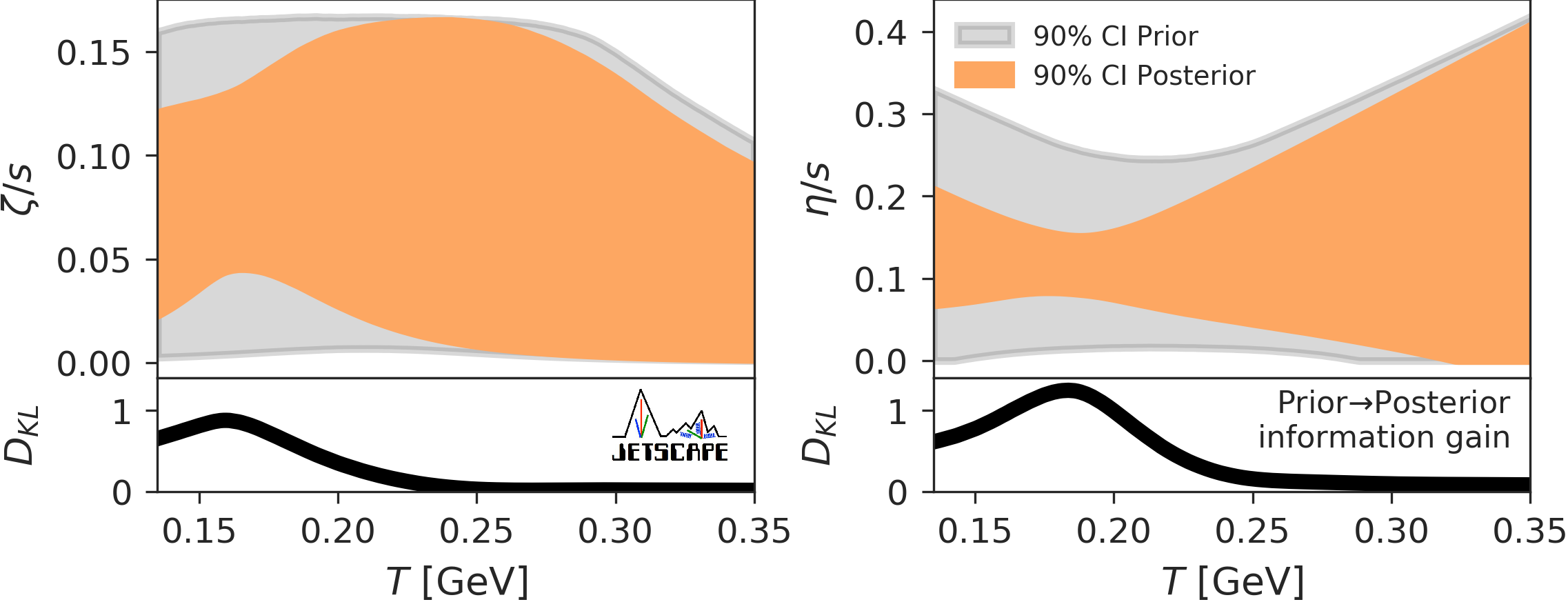}
\caption{$90$\% credible intervals for the priors (gray) and Bayesian model averaged posteriors for the specific bulk (left) and shear (right) viscosities, along with their corresponding information gain (Kullback-Leibler divergence $D_{KL}$).}
\label{F3}
\end{figure}

To emphasize the constraints provided by the experimental data, we calculate the information gain of our posteriors for the temperature dependence of the viscosities of QCD, relative to the corresponding priors, using the Kullback-Leibler divergence ($D_{KL}$) \cite{kullback1951}. We show the result in Fig.\,\ref{F3} alongside the 90\% prior and Bayesian model averaged posteriors. While the experimental data are seen to provide significant constraints for $150\lesssim T\lesssim 250$\,MeV their constraining power rapidly degrades at higher temperatures. In the deconfinement region, the most likely values for $\eta/s$ are of order $0.1$; $\zeta/s$ also favors values around $0.05{-}0.1$ in that region, although constraints are weaker than for $\eta/s$. Stronger priors could be used to further constrain the favored viscosity values: for example, negative slopes for the shear viscosity at high temperature could be excluded based on theoretical guidance \cite{Bernhard:2019bmu}. We elect not to do so, emphasizing instead the constraining power provided directly by measurements.

\noindent \textbf{\textit{Summary.}}
%
Theoretical insights into the transport properties of quark-gluon plasma are still limited for temperatures ${\sim\,}150{-}350$\,MeV. The phenomenological constraints obtained in this work from heavy-ion measurements complement the current theoretical knowledge, supplementing a range of calculations of the shear and bulk viscosities of the quark-gluon plasma at lower~\cite{Lu:2011df, Rose:2017bjz, Rose:2020lfc}, intermediate \cite{Karsch:2007jc, NoronhaHostler:2008ju, Bazavov:2019lgz} and higher \cite{Arnold:2006fz, Ghiglieri:2018dib} temperatures.

In this work, we obtained new state-of-the-art estimates for the QGP shear and bulk viscosities with more robust estimates for the uncertainties of these key transport coefficients. We introduced Model Averaging into Bayesian inference to include both experimental and known theoretical uncertainties in the uncertainty budget for the model parameters inferred from RHIC and LHC data. By allowing for a systematic inclusion of (i) additional measurements, (ii) model uncertainties, (iii) error correlations, and (iv) more rigorous and objective specification of model priors, the methods pioneered in this analysis for heavy-ion physics provide a clear path forward for rigorous estimations of the transport properties of the quark-gluon plasma.


\noindent{\it Acknowledgments:} 
We thank Jonah Bernhard, Gabriel Denicol, Scott Moreland, Scott Pratt and Derek Teaney for useful discussions, and Richard J. Furnstahl and Xilin Zhang for their insights on Bayesian Inference and Markov Chain Monte Carlo. This work was supported in part by the National Science Foundation (NSF) within the framework of the JETSCAPE collaboration, under grant numbers ACI-1550172 (Y.C. and G.R.), ACI-1550221 (R.J.F., F.G., M.K. and B.K.), ACI-1550223 (D.E., M.M., U.H., L.D., and D.L.), ACI-1550225 (S.A.B., J.C., T.D., W.F., W.K., R.W., S.M., and Y.X.), ACI-1550228 (J.M., B.J., P.J., X.-N.W.), and ACI-1550300 (S.C., L.C., A.K., A.M., C.N., A.S., J.P., L.S., C.Si., R.A.S. and G.V.). It was also supported in part by the NSF under grant numbersPHY-1516590, 1812431 and PHY-2012922 (R.J.F., B.K., F.G., M.K., and C.S.), and by the US Department of Energy, Office of Science, Office of Nuclear Physics under grant numbers \rm{DE-AC02-05CH11231} (D.O., X.-N.W.), \rm{DE-AC52-07NA27344} (A.A., R.A.S.), \rm{DE-SC0013460} (S.C., A.K., A.M., C.S. and C.Si.), \rm{DE-SC0004286} (L.D., M.M., D.E., U.H. and D.L.), \rm{DE-SC0012704} (B.S. and C.S.), \rm{DE-FG02-92ER40713} (J.P.) and \rm{DE-FG02-05ER41367} (T.D., W.K., J.-F.P., S.A.B. and Y.X.). The work was also supported in part by the National Science Foundation of China (NSFC) under grant numbers 11935007, 11861131009 and 11890714 (Y.H. and X.-N.W.), by the Natural Sciences and Engineering Research Council of Canada (C.G., M.H., S.J., C.P. and G.V.), by the Fonds de Recherche du Qu\'{e}bec Nature et Technologies (FRQ-NT) (G.V.), by the Office of the Vice President for Research (OVPR) at Wayne State University (Y.T.), by the S\~{a}o Paulo Research Foundation (FAPESP) under projects 2016/24029-6, 2017/05685-2 and 2018/24720-6 (M.L.), and by the University of California, Berkeley - Central China Normal University Collaboration Grant (W.K.). U.H. would like to acknowledge support by the Alexander von Humboldt Foundation through a Humboldt Research Award. Allocation of supercomputing resources (Project: PHY180035) were obtained in part through the Extreme Science and Engineering Discovery Environment (XSEDE), which is supported by National Science Foundation grant number ACI-1548562. Calculation were performed in part on Stampede2 compute nodes, generously funded by the National Science Foundation (NSF) through award ACI-1134872, within the Texas Advanced Computing Center (TACC) at the University of Texas at Austin \cite{TACC}, and in part on the Ohio Supercomputer \cite{OhioSupercomputerCenter1987} (Project PAS0254). Computations were also carried out on the Wayne State Grid funded by the Wayne State OVPR, and on the supercomputer \emph{Guillimin} from McGill University, managed by Calcul Qu\'{e}bec and Compute Canada. The operation of the supercomputer \emph{Guillimin} is funded by the Canada Foundation for Innovation (CFI), NanoQu\'{e}bec, R\'{e}seau de M\'{e}dicine G\'{e}n\'{e}tique Appliqu\'{e}e~(RMGA) and FRQ-NT.  Data storage was provided in part by the OSIRIS project supported by the National Science Foundation under grant number OAC-1541335.

\bibliography{biblio_inspire,biblio_other}

\end{document}